\documentclass[conference]{IEEEtran}
\IEEEoverridecommandlockouts
\usepackage{cite}
\usepackage{amsmath,amssymb,amsfonts}
\usepackage{graphicx}
\usepackage{textcomp}
\usepackage{xcolor}
\def\BibTeX{{\rm B\kern-.05em{\sc i\kern-.025em b}\kern-.08em
    T\kern-.1667em\lower.7ex\hbox{E}\kern-.125emX}}
\usepackage{algorithm}
\usepackage{algpseudocode}

\begin{document}

\title{Multi Objective Resource Optimization of Wireless Network Based on Cross Domain Virtual Network Embedding
\thanks{This work is partially supported by the Open Research Fund of State Key Laboratory of Space-Ground Integrated Information Technology under grant NO.2018\_SGIIT\_KFJJ\_TX\_03, partially supported by the Major Scientific and Technological Projects of CNPC under Grant ZD2019-183-006, and partially supported by ``the Fundamental Research Funds for the Central Universities" of China University of Petroleum (East China) under Grant 20CX05017A. (Corresponding author: Tao Dong.) }
}

\author{Chao Wang\IEEEauthorrefmark{1}, Tao Dong\IEEEauthorrefmark{2}, Youxiang Duan\IEEEauthorrefmark{3}, Qifeng Sun\IEEEauthorrefmark{4}, and Peiying Zhang\IEEEauthorrefmark{5} \\

\IEEEauthorblockA{\IEEEauthorrefmark{1}\IEEEauthorrefmark{3}\IEEEauthorrefmark{4}\IEEEauthorrefmark{5} College of Computer Science and Technology, China University of Petroleum (East China), Qingdao, 266580, China.}
\IEEEauthorblockA{\IEEEauthorrefmark{2} State Key Laboratory of Space-Ground Integrated Information Technology, \\ Beijing Institute of Satellite Information Engineering, Beijing, 100000, China.}
(e-mail: wangch\_upc@qq.com, dongtaoandy@163.com, yxduan@upc.edu.cn, sunqf@upc.edu.cn, zhangpeiying@upc.edu.cn)}

\maketitle

\begin{abstract}
The rapid development of virtual network architecture makes it possible for wireless network to be widely used. With the popularity of artificial intelligence (AI) industry in daily life, efficient resource allocation of wireless network has become a problem. Especially when network users request wireless network resources from different management domains, they still face many practical problems. From the perspective of virtual network embedding (VNE), this paper designs and implements a multi-objective optimization VNE algorithm for wireless network resource allocation. Resource allocation in virtual network is essentially a problem of allocating underlying resources for virtual network requests (VNRs). According to the proposed objective formula, we consider the optimization mapping cost, network delay and VNR acceptance rate. VNE is completed by node mapping and link mapping. In the experiment and simulation stage, it is compared with other VNE algorithms, the cross domain VNE algorithm proposed in this paper is optimal in the above three indicators. This shows the effectiveness of the algorithm in wireless network resource allocation.
\end{abstract}


\section{Introduction}\label{part1}

\begin{figure*}[htbp]
\centering
\includegraphics[width=0.75\textwidth]{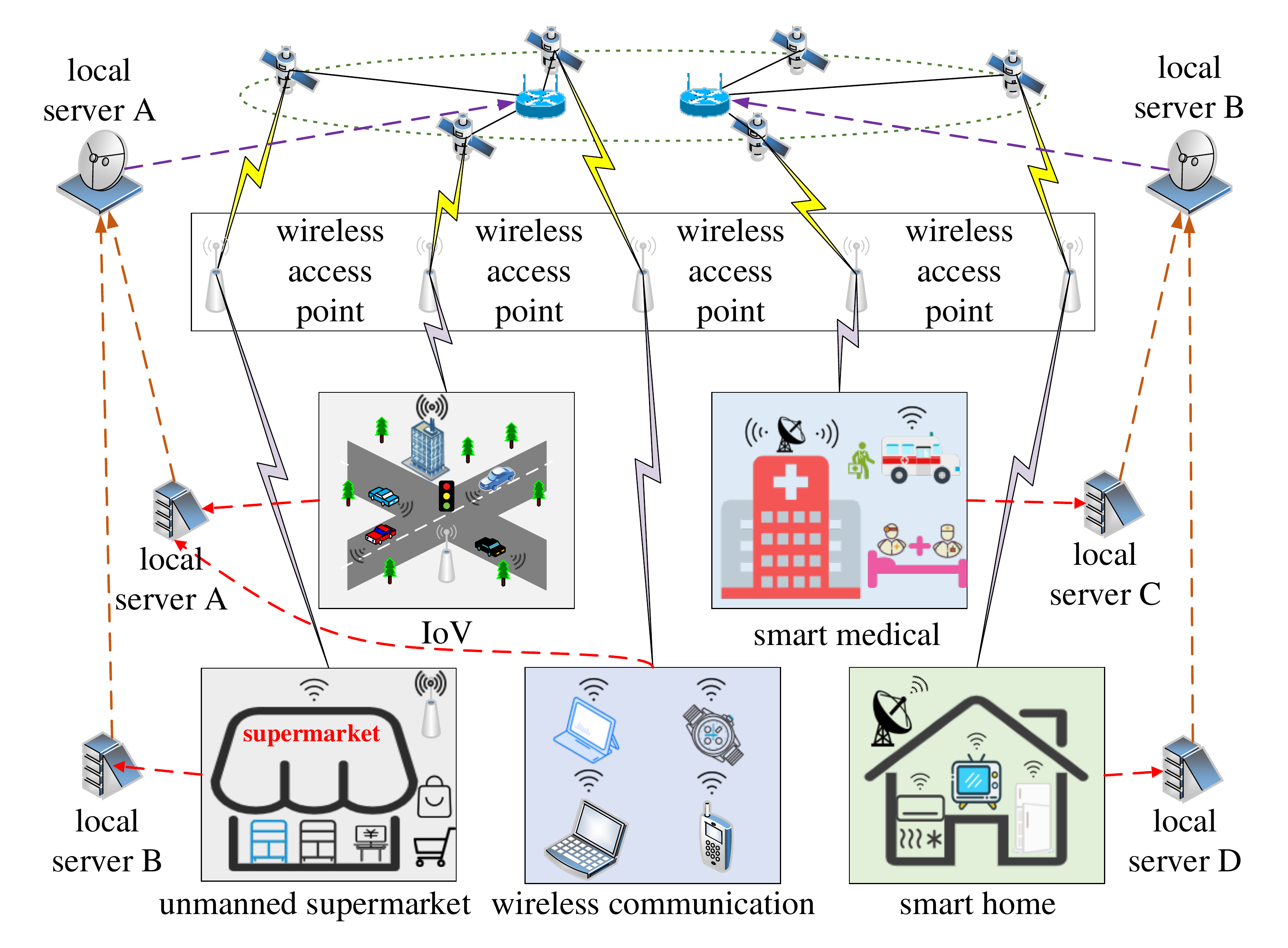}
\caption{Application scenarios of wireless network.}
\label{fig_1}
\end{figure*}

In order to serve the rapid development needs of modern society, modern network service models have formed a situation of coordinated services in multiple heterogeneous networks such as fixed networks, mobile networks and space networks \cite{j1,z1}. Wireless network is a typical representative of the above-mentioned networks and has been extremely widely used by virtue of its flexibility of not requiring wired connections \cite{j2}. With the rapid development of network virtualization (NV) technology, virtual network architecture has gradually begun to serve as an infrastructure to support the development of wireless networks \cite{r3,z4,r4}. As a basic network, wireless networks are widely used in a variety of artificial intelligence (AI) scenarios. Its application scenario is shown in Fig. \ref{fig_1}. In the virtual network environment, more and more wireless terminal users access the wireless network, they constantly request resources to the wireless network center, so the wireless network center is facing a huge pressure of resource allocation. Radio network resource management faces severe challenges, including storage, spectrum, computing resource allocation, and joint allocation of multiple resources \cite{jcx1,jcx2}. With the rapid development of communication networks, the integrated space-ground network has also become a key research object \cite{jcx3}.

With the continuous exploration and innovation of network technology, NV technology is considered to be a promising application technology that can be used as the infrastructure of future networks. As one of the most concerned issues of NV, the essential problem to be solved by virtual network embedding (VNE) technology is the resource allocation problem of the underlying network \cite{z2,r1}. Under the virtual network architecture, the process of end users requesting resources from the wireless network is the embedding process of virtual network requests (VNRs). Currently, a complete wireless network covering the world has not been formed. Due to factors such as geographic location, wireless networks in different regions may belong to different network service providers (SPs) \cite{z5}. There are commercial competition and other relationships among SPs, and they are unwilling to disclose user information and network topology information in their domains, which brings difficulties to the allocation of underlying network resources \cite{r2}. Cross-domain VNE technology is a reliable way to solve the above problems.

Since the application of wireless networks, scholars and researchers have fully explored the resource allocation structure and allocation methods under this network. Xu et al. \cite{b1} proposed a dynamic resource allocation method based on load balancing and QoS. The main purpose of this method was to solve the problem of solidifying the optical fiber wireless access (FiWi) network. This method was divided into three sub directions: (1) In order to describe the real-time consumption of the underlying network resources, a time window based update mechanism of the underlying network resources was implemented. (2) In order to accurately calculated the priority of VNRs, a QoS based VNR ranking mechanism was proposed, which can reasonably sort VNRs. (3) A resource allocation mechanism based on load balancing was implemented to avoid unbalanced resource consumption. The authors took into account the accuracy and balance of wireless network resource allocation. They were outstanding representatives in the field of wireless network resource allocation.

Miao et al. \cite{b2} studied the joint configuration of communication, computing and popular content resources in wireless networks from the perspective of mobile virtual network operators (MVNOs). The authors summarized the MVNO problem as a distribution game and by adopting a supply-demand game model, the optimal solution for matching between contents, computation nodes, MVNOs, and users was given. So the authors focused on the issue of MVNO revenue. Also in reference \cite{b3,b4} different wireless network resource allocation strategies were proposed. The former focused on the issue of cooperative resource allocation in multi-carrier cognitive radio (CR) networks. Under the condition of minimizing the rate of the primary user (PU), the resource allocation process of maximizing the rate of the secondary user (SU) was realized. The latter mainly focused on the multi-user resource allocation problem of the wireless uplink cellular network. In order to cope with the limited total transmission time of relay nodes and limited available energy, the authors designed and implemented two different relay-based harvest-retransmission scenarios. The results proved that the above strategies had achieved good results and related work has also been done \cite{b5,b6} and so on.

The main work of this paper is different from the above research, mainly reflected in the design and implementation of a low-level VNE algorithm, not only for a specific wireless network scene. Specifically, the main work of this paper is as follows.

\begin{itemize}

\item We propose a multi-objective optimization VNE algorithm to solve the resource allocation problem in wireless networks. The algorithm not only optimizes a single network resource, but also optimizes the embedding cost, network delay and VNR acceptance rate.
\item This paper studies the VNE problem across multi-domain wireless networks, relying on the cooperation of local controllers and global controllers to complete the rational allocation of underlying resources.
\item In the simulation stage, we design experiments to verify the performance of the multi-objective algorithm. We compare with several other algorithms in VNE cost, network delay and VNR acceptance rate. The result proves that the algorithm proposed in this paper has obvious advantages.

\end{itemize}

Paper organization structure: In section \ref{part1}, we introduce the basic concepts of wireless network and cross domain VNE, which leads to the problem of wireless network resource allocation. In section \ref{part2}, we describe the related problems of cross-domain virtual network and establish the necessary network model. Section \ref{part3} introduces the constraints of cross domain VNE and the evaluation indexes used in the experimental comparison phase. Section \ref{part4} introduces the implementation process of the algorithm. Section \ref{part5} shows and analyzes the experimental results. The last section summarizes the whole paper.

\section{Problem Description and Network Model}\label{part2}

\subsection{Description of Cross-Domain Virtual Network Embedding Problem}

In the future virtual network architecture, the network will be divided into autonomous domains. The region of each network autonomous region is different from its administrative organization, and because of the existence of commercial competition and other factors, the network topology information in each autonomous region is not open to the public \cite{z6,j3}. In this case, the research of cross domain VNE algorithm is more practical than single domain VNE algorithm \cite{z3}.

Based on the virtual network as the infrastructure, wireless network can be more widely used. The resource request from users accessing wireless network is VNR. VNR usually consists of virtual node and virtual link. The virtual node requests the node resources (CPU, storage space, etc.) of the underlying network, and the virtual link requests the link resources (bandwidth, delay, etc.) of the underlying network. Since VNRs may require wireless network resources in different regions, these wireless networks may be managed by different managers. So at this time, different virtual nodes in the same VNR may be mapped to multiple physical domains, and the virtual links between virtual nodes may also be mapped to different physical links. This is the problem of cross domain VNE.

\subsection{Network Model}

The physical network is modeled as an undirected weighted graph represented by $G_i^S=\{N_i^S,L_i^S,A_{N_i}^S,A_{L_i}^S\}$, where $A_{N_i}^S=\{CPU_{N_i}^S,P_{N_i}^S,D_{N_i}^S\}$, $A_{L_i}^S=\{BW_{L_i}^S,P_{L_i}^S,D_{L_i}^S\}$. $G_i^S$ represents the $i$-th physical domain. $N_i^S$ represents the set of all physical nodes in the $i$-th physical domain. $L_i^S$ represents the set of all physical links in the $i$-th physical domain. $A_{N_i}^S$ represents the attribute set of physical node in the $i$-th physical domain, including CPU resource $CPU_{N_i}^S$, CPU resource unit price $P_{N_i}^S$ and delay $D_{N_i}^S$. $A_{L_i}^S$ represents the attribute set of physical links in the $i$-th physical domain, including bandwidth resource $BW_{L_i}^S$, bandwidth resource unit price $P_{L_i}^S$ and delay $D_{L_i}^S$. Links between physical domains also include the above attributes.

The virtual network is modeled as an undirected weighted graph represented by $G_i^V=\{N_i^V,L_i^V,CPU_{N_i}^V,BW_{L_i}^V\}$. Where $G_i^V$ is the $i$-th VNR, $N_i^V$ is the set of virtual nodes in the $i$-th VNR, corresponding to the CPU resource requirement $CPU_{N_i}^V$ of the virtual node. $L_i^V$ represents the set of virtual links in the $i$-th VNR, corresponding to the bandwidth resource requirement $BW_{L_i}^V$ of the virtual link.

\begin{figure}[htbp]
\centering
\includegraphics[width=0.95\columnwidth]{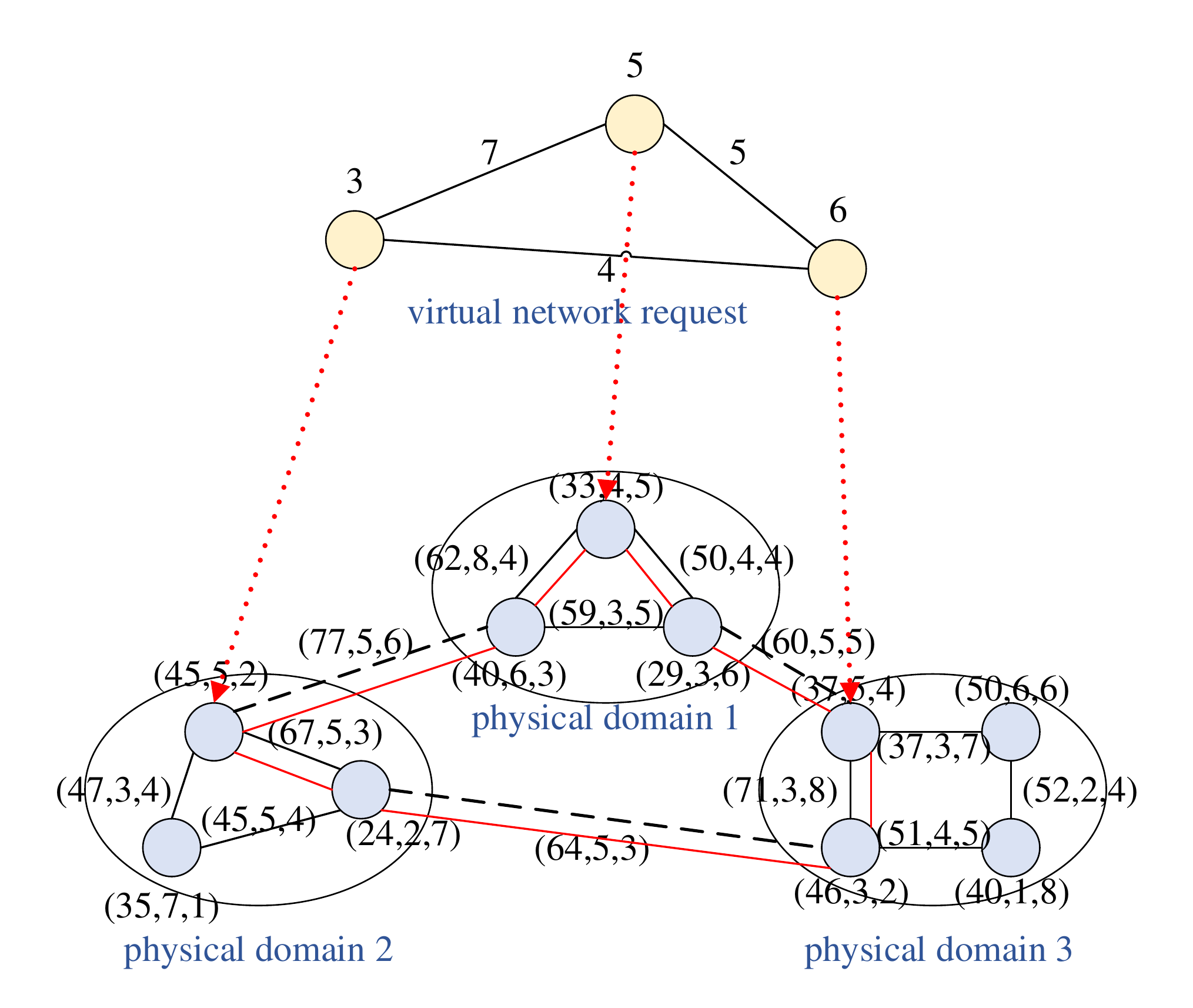}
\caption{Schematic diagram of virtual network embedded in physical network.}
\label{fig_2}
\end{figure}

Fig. \ref{fig_2} shows the embedding of a VNR to a cross-domain physical network. The numbers next to nodes and links correspond to the attributes described earlier. For physical nodes, the numbers in brackets represent the amount of CPU resources, the unit price of CPU resources and the node delay in turn. For physical links, the numbers in brackets represent the amount of bandwidth resources, the unit price of bandwidth resources and the link delay in turn.

\section{Constraints and Evaluation Indicators}\label{part3}

\subsection{Constraints}

Each VNR consumes a certain amount of physical network resources. When the physical network resources are insufficient, the efficiency of VNE will be affected. Therefore, VNE needs to follow some constraints (node constraints and link constraints).

\textit{Node constraints:}

\begin{equation}
n^v \in N_i^V, \, n^s \in N_i^S \,\, or \,\, n^s \in N_j^S, \, if \, n^v \uparrow n^s.\label{eq1}
\end{equation}

\eqref{eq1} indicates that if the target physical node of virtual node $n^v$ is $n^s$, then the physical domain to which physical node $n^s$ belongs is the candidate domain of virtual node $n^v$. That is, virtual nodes can only be mapped to candidate physical domains.

\begin{equation}
CPU(n^v) \leq CPU(n^s), \, if \, n^v \uparrow n^s.\label{eq2}
\end{equation}

\eqref{eq2} indicates that if the target physical node of virtual node $n^v$ is $n^s$, then physical node $n^s$ needs to meet the CPU resource requirements of virtual node $n^v$.

\textit{Link constraints:}

\begin{equation}
BW(l^v) \leq BW(l^s), \, if \, l^v \uparrow l^s.\label{eq3}
\end{equation}

\eqref{eq3} indicates that if the target physical link of virtual link $l^v$ is $l^s$, then physical link $l^s$ needs to meet the bandwidth resource requirements of virtual link $l^v$.

\begin{equation}
l^v \uparrow (l_1^s \cup l_2^s \cup ... \cup l_k^s).\label{eq4}
\end{equation}

\eqref{eq4} indicates that a virtual link $l^v$ may be mapped to multiple physical links (path splitting).

\subsection{Evaluation Indicators}

In this paper, a multi-objective optimization algorithm for VNE is proposed, which takes into account many indicators of VNE, including the cost of VNE, the overall delay and the VNR acceptance rate. In order to get the optimal algorithm, the overall objective function is as follows:

\begin{equation}
\begin{aligned}
min[OBJ=\sum_{all \, n^v,n^v \uparrow n^s}(CPU(n^s) \times P(n^s) + D(n^s)) \\ + \sum_{all \, l^v \uparrow l^s}(BW(l^s) \times P(l^s)+D(l^s))].\label{eq5}
\end{aligned}
\end{equation}

The objective function takes into account the embedding cost and delay of virtual network.

The cost of a successful VNE is as follows:

\begin{equation}
C=\sum_{n^v \in N_i^V}CPU(n^v)+\sum_{l^v \in L_i^V}BW(l^v) \times num\_{hops(l^v)}.\label{eq6}
\end{equation}

The embedded cost of virtual network is determined by the consumption of CPU resources and bandwidth resources, where $num\_hops(l^v)$ is the hop number of virtual link $l^v$ where path splitting occurs.

The VNR acceptance rate is calculated as follows:

\begin{equation}
acc=\lim_{T \to \infty} \frac{\sum\limits_t^TVNR_{acc}}{\sum\limits_t^TVNR_{arr}}.\label{eq7}
\end{equation}

In \eqref{eq7}, $\sum\limits_t^TVNR_{acc}$ represents the number of VNRs that are completely embedded in the underlying network, that is, resources are successfully allocated for the VNR. $\sum\limits_t^TVNR_{arr}$ represents the total number of VNRs that come within the time range $t$.

\section{Algorithm Description and Implementation}\label{part4}

The multi-objective resource optimization algorithm based on cross domain VNE adopts a centralized multi-layer VNE architecture. Specifically, different wireless network domains are managed by different local controllers, which manage the specific network topology information in the domain \cite{z7}. All local controllers are managed by a global controller. The global controller is responsible for the overall management of each local domain information and is responsible for the mapping of inter-domain links. The overall flow of the algorithm is as follows:

\begin{itemize}

\item The resource request sent by the wireless network user first reaches the global controller, then the global controller divides it.
\item The global controller sends the VNR subgraph to each local controller. The local controller derives candidate nodes based on the subgraph information and the topology resource information in the domain.
\item The local controller uploads the set of candidate nodes and the link information of each virtual node to the global controller. According to the objective function, the VNR is pre mapped under the operation of the global controller.
\item After receiving the pre mapping results, the local controller maps the VNR according to the topology resource information in the domain.
\item The global controller receives the intra domain mapping results from the local controller and performs inter domain link mapping. If all nodes and links of a VNR are mapped successfully, the whole mapping process is completed.

\end{itemize}

The pseudo code for the two key processes of candidate node selection and intra-domain mapping is given below.

\subsection{Candidate Node Selection}

According to the \eqref{eq6}, each virtual node is selected from its candidate domain. The candidate physical nodes are sorted according to the mapping cost, and the physical nodes with the least mapping cost and not mapped are selected to map.

\begin{algorithm}
  \caption{Candidate Node Selection}
  \begin{algorithmic}[1]
    \Require
        {$G_i^S$}, {$G_i^V$};
    \Ensure
        {$Node\_set$};
    \For {$all \, n^s \in G_i^S$}
    \State $C=\sum\limits_{n^v \in N_i^V}CPU(n^v)+\sum\limits_{l^v \in L_i^V}BW(l^v) \times num\_{hops(l^v)}$;
    \EndFor
    \State $sort(MappingCost)$;
    \While {$n^s \in G_i^S \, \& \, n^s \in Candi\_Domain$}
    \State $n_{Candi}^s=getMinCostNode()$;
    \If {$isMarked(n_{Candi}^s)\,\, is\,\,FALSE$}
    \State $setMarked(TRUE)$;
    \State $setCandiNode(TRUE)$;
    \EndIf
    \EndWhile
    \State \Return $CandiNode$;
  \end{algorithmic}
\end{algorithm}

\subsection{Intra Domain Mapping}

The VNE process in each physical domain is carried out under the control of local controller. The virtual nodes in the domain are mapped according to the previous candidate node results. After that, Floyd algorithm is used to map the intra domain links.

\begin{algorithm}
  \caption{Intra Domain Mapping}
  \begin{algorithmic}[1]
     \Require
        $CandiNode$, {$G_i^S$}, {$G_i^V$};
    \Ensure
        ${Intra\,Domain\,Mapping\,Scheme}$;
    \For {$n^s \in CandiNode$}
    \State $addPhysicalNodeToRequest(n^s)$;
    \EndFor
    \For {$n_1^v \in G_i^V$}
    \For {$n_2^v \in G_i^V$}
    \If {$VirtualLink(n_1^v,n_2^v)\,\,is\,\,TRUE$}
    \State $addPhysicalLinkToRequest()$;
    \EndIf
    \EndFor
    \EndFor
    \State \Return $Intra\,Domain\,Mapping\,Scheme$;
  \end{algorithmic}
\end{algorithm}

\subsection{Algorithm Complexity}

Suppose that the number of all virtual nodes is $n$ in the process of VNR arrival. The number of physical nodes in all physical domains is $m$. The number of physical domains is represented by $d$. Then the overall complexity of the algorithm is:

\begin{equation}
O(n,m,d)=O(n \cdot m \cdot n \cdot d)=O(n^2 \cdot m \cdot d).\label{eq8}
\end{equation}

In the VNR process, $m$ and $d$ are fixed values, so the overall complexity of the algorithm is $O(n^2)$.

\section{Experimental Design and Results}\label{part5}

\subsection{Parameter Setting}

The parameter information to be set mainly includes physical network (node, link) attribute and virtual network request (node, link) attribute. The specific parameter information is shown in TABLE \ref{tab1}.

\begin{table}[htbp]
\caption{Parameter Setting}
\begin{center}
\begin{tabular}{|p{40mm} p{20mm}|}
\hline
Parameter name & Parameter value  \\
\hline
\multicolumn{2}{|c|}{\textbf{Physical Network Attributes}} \\
\hline
physical domain & 4 \\
\hline
CPU resource & U[100,300] \\
\hline
physical nodes / domain & 30 \\
\hline
node delay & U[1,10] \\
\hline
CPU cost & U[1,10] \\
\hline
bandwidth resource & U[1000,3000] \\
\hline
link delay & U[1,10] \\
\hline
bandwidth cost & U[1,10] \\
\hline
\multicolumn{2}{|c|}{\textbf{Virtual Network Attributes}} \\
\hline
virtual nodes & U[2,6] \\
\hline
CPU requirements & U[1,10] \\
\hline
bandwidth requirements & U[5,15] \\
\hline
\end{tabular}
\label{tab1}
\end{center}
\end{table}

\subsection{Experimental Results and Analysis}

In the experimental simulation stage, we compare the multi-objective optimization algorithm based on cross domain VNE (MOO-VNE) with other representative cross domain VNE algorithms in three aspects of VNE cost, network delay, and VNR acceptance rate to verify the actual performance of the multi-objective optimization algorithm. TABLE \ref{tab2} shows the comparison of several algorithms.

\begin{table}[htbp]
\caption{Algorithm Ideas}
\begin{center}
\begin{tabular}{|p{20mm}|p{60mm}|}
\hline
Name & Concept \\
\hline
MOO-VNE & According to the mapping cost formula, candidate nodes are selected in advance, then Floyd algorithm is used to complete the link mapping. \\
\hline
PSO-VNE  \cite{b7} & The candidate nodes are selected based on the number of hops from the boundary node. Use particle swarm optimization algorithm to optimize the node mapping scheme. \\
\hline
MC-VNE \cite{b8} & The idea of Kruskal spanning tree is used for link mapping first, and the node mapping scheme is determined by the link mapping scheme. \\
\hline
\end{tabular}
\label{tab2}
\end{center}
\end{table}

\begin{figure*}[htbp]
\centering
\includegraphics[width=1\textwidth]{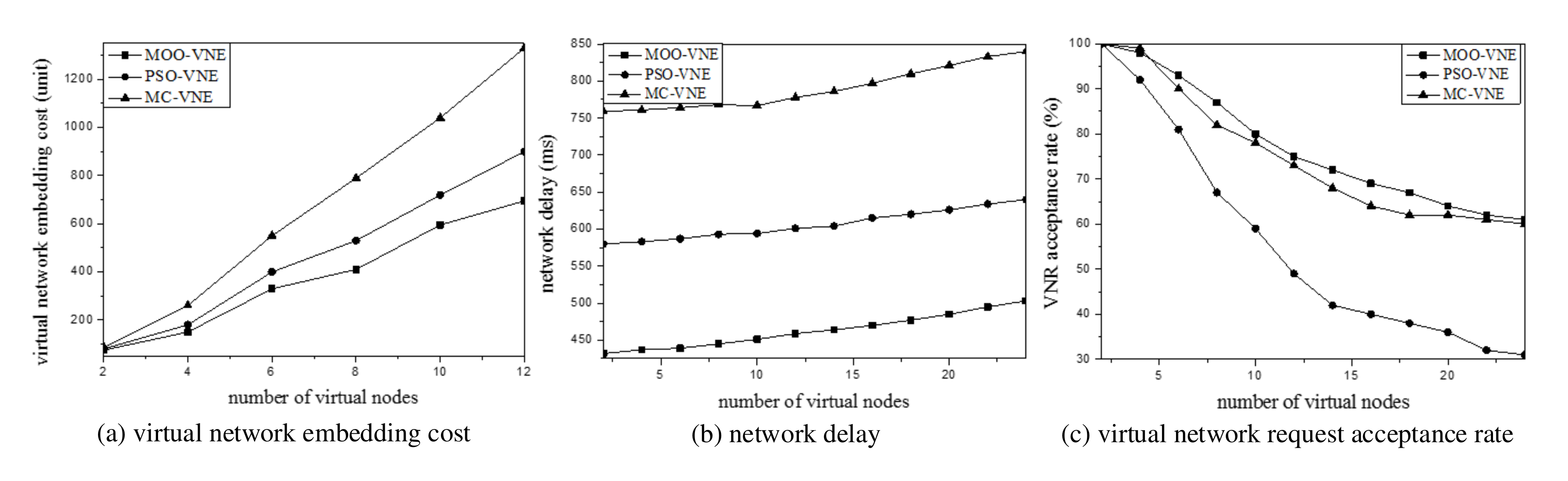}
\caption{Experimental results. (a) Virtual network embedding cost. (b) Network delay. (c) Virtual network request acceptance rate.}
\label{fig_3}
\end{figure*}

The comparison of experimental results of VNE cost, network delay, and VNR acceptance rate are shown in Fig. \ref{fig_3}. Overall, the performance of MOO-VNE algorithm is optimal in the above three aspects.

In terms of VNE cost and network delay, the MOO-VNE algorithm takes into account the two indicators of cost and delay when selecting candidate nodes. The two physical nodes with the smallest indicators are comprehensively selected as candidate nodes, so the experimental results are better in these two aspects. In terms of the acceptance rate of VNRs, the MOO-VNE algorithm takes into account the consumption of node resources and link resources, and uses the idea of load balancing to make the algorithm allocate physical network resources reasonably in order to satisfy more VNRs. The disadvantage of the PSO-VNE algorithm is that it always prioritizes the boundary node as the candidate node and selects the candidate node by the number of hops from the boundary node, which limits the possibility of the optimal solution, so the overall performance is poor. The disadvantage of the MC-VNM algorithm is that it adopts a greedy strategy. The effect of relying on Kruskal's spanning tree optimization method is obviously not as good as that of the heuristic algorithm. Therefore, the comprehensive performance of the MOO-VNE algorithm shown above is optimal, and the effect of multi-objective optimization is achieved.

\section{Conclusion}\label{part6}

The problem of resource allocation in wireless networks is a difficult problem, especially when facing user resource requests from different physical domains. We model the resource allocation problem of the wireless network as a cross-domain VNE problem. In order to allocate network resources reasonably, we design and realize a multi-objective optimization the cross-domain VNE algorithm is optimized in three aspects: VNE cost, network delay, and VNR acceptance rate. We select candidate nodes based on the objective function, then use the Floyd algorithm to complete the link mapping. In the experimental stage, by comparing with several other representative algorithms, it proved the excellent performance of MOO-VNE algorithm in VNE cost, network delay and VNR acceptance rate. Therefore, the MOO-VNE algorithm has a certain enlightening effect on the resource allocation of wireless networks.

\section*{Acknowledgment}

This work is partially supported by the Open Research Fund of State Key Laboratory of Space-Ground Integrated Information Technology under grant NO.2018\_SGIIT\_KFJJ\_TX\_03, partially supported by the Major Scientific and Technological Projects of CNPC under Grant ZD2019-183-006, and partially supported by ``the Fundamental Research Funds for the Central Universities" of China University of Petroleum (East China) under Grant 20CX05017A.

\vspace{12pt}


\begin{thebibliography}{00}

\bibitem{j1}
J. Du, C. Jiang, Z. Han, H. Zhang, S. Mumtaz and Y. Ren, “Contract Mechanism and Performance Analysis for Data Transaction in Mobile Social Networks,” IEEE Transactions on Network Science and Engineering, vol. 6, no. 2, pp. 103-115, 2019.

\bibitem{z1}
P. Zhang, H. Yao, M. Li and Y. Liu, “Virtual network embedding based on modified genetic algorithm,” Peer-to-Peer Networking and Applications, vol. 2, pp. 1-12, 2017.

\bibitem{j2}
C. Jiang, H. Zhang, Y. Ren, Z. Han, K. Chen and L. Hanzo, “Machine Learning Paradigms for Next-Generation Wireless Networks,” IEEE Wireless Communications, vol. 24, no. 2, pp. 98-105, 2017.

\bibitem{r3}
M. G. Kibria, K. Nguyen, G. P. Villardi, O. Zhao, K. Ishizu and F. Kojima, “Big Data Analytics, Machine Learning, and Artificial Intelligence in Next-Generation Wireless Networks,” IEEE Access, vol. 6, pp. 32328-32338, 2018.

\bibitem{z4}
P. Zhang, X. Pang, N. Kumar, G. S. Aujla and H. Cao, “A Reliable Data-Transmission Mechanism Using Blockchain in Edge Computing Scenarios,” IEEE Internet of Things Journal, doi: 10.1109/JIOT.2020.3021457, pp. 1-1, 2020.

\bibitem{r4}
S. Xu et al., “Load-Balancing and QoS Based Dynamic Resource Allocation Method for Smart Gird Fiber-Wireless Networks,” Chinese Journal of Electronics, vol. 28, no. 6, pp. 1234-1243, 2019.

\bibitem{jcx1}
C. Jiang, Y. Chen, K. J. R. Liu and Y. Ren, ``Renewal-Theoretical Dynamic Spectrum Access in Cognitive Radio Network with Unknown Primary Behavior,'' {\em IEEE Journal on Selected Areas in Communications}, vol. 31, no. 3, pp. 406-416, 2013.

\bibitem{jcx2}
C. Jiang, Y. Chen, Y. Gao and K. J. R. Liu, ``Joint Spectrum Sensing and Access Evolutionary Game in Cognitive Radio Networks,'' {\em IEEE Transactions on Wireless Communications}, vol. 12, no. 5, pp. 2470-2483, 2013.

\bibitem{jcx3}
X. Zhu, C. Jiang, L. Kuang, N. Ge and J. Lu, ``Non-orthogonal Multiple Access Based Integrated Terrestrial-Satellite Networks,'' {\em IEEE Journal on Selected Areas in Communications}, vol. 35, no. 10, pp. 2253-2267, Oct. 2017.

\bibitem{z2}
P. Zhang, H. Yao and Y. Liu, “Virtual Network Embedding Based on Computing, Network, and Storage Resource Constraints,” IEEE Internet of Things Journal, vol. 5, no. 5, pp. 3298-3304, 2018.

\bibitem{r1}
Y. Ni, G. Huang, S. Wu, C. Li, P. Zhang and H. Yao, “A PSO based multi-domain virtual network embedding approach,” China Communications, vol. 16, no. 4, pp. 105-119, 2019.

\bibitem{z5}
P. Zhang, C. Jiang, X. Pang and Y. Qian, “STEC-IoT: A Security Tactic by Virtualizing Edge Computing on IoT,”  IEEE Internet of Things Journal, doi: 10.1109/JIOT.2020.3017742, pp. 1-1, 2020.

\bibitem{r2}
H. Cao, Y. Guo, Y. Hu, S. Wu, H. Zhu and L. Yang, “Location Aware and Node Ranking Value-Assisted Embedding Algorithm for One-Stage Embedding in Multiple Distributed Virtual Network Embedding,” IEEE Access, vol. 6, pp. 78425-78436, 2018.

\bibitem{b1}
S. Xu et al., ``Load-Balancing and QoS Based Dynamic Resource Allocation Method for Smart Gird Fiber-Wireless Networks,'' Chinese Journal of Electronics, vol. 28, no. 6, pp. 1234-1243, Nov. 2019.

\bibitem{b2}
Z. Miao, Y. Wang and Z. Han, ``A Supplier-Firm-Buyer Framework for Computation and Content Resource Assignment in Wireless Virtual Networks,'' IEEE Transactions on Wireless Communications, vol. 18, no. 8, pp. 4116-4128, Aug. 2019.

\bibitem{b3}
D. Xu and Q. Li, ``Cooperative Resource Allocation in Cognitive Radio Networks With Wireless Powered Primary Users,'' IEEE Wireless Communications Letters, vol. 6, no. 5, pp. 658-661, Oct. 2017.

\bibitem{b4}
S. Lohani, R. A. Loodaricheh, E. Hossain and V. K. Bhargava, ``On Multiuser Resource Allocation in Relay-Based Wireless-Powered Uplink Cellular Networks,''  IEEE Transactions on Wireless Communications, vol. 15, no. 3, pp. 1851-1865, March 2016.

\bibitem{b5}
L. Tan, Z. Zhu, F. Ge and N. Xiong, ``Utility Maximization Resource Allocation in Wireless Networks: Methods and Algorithms,'' IEEE Transactions on Systems, Man, and Cybernetics: Systems, vol. 45, no. 7, pp. 1018-1034, July 2015.

\bibitem{b6}
S. Xu, P. Li, S. Guo and X. Qiu, ``Fiber-Wireless Network Virtual Resource Embedding Method Based on Load Balancing and Priority,'' IEEE Access, vol. 6, pp. 33201-33215, 2018.

\bibitem{z6}
P. Zhang, C. Wang, C. Jiang and A. Benslimane, “Security-Aware Virtual Network Embedding Algorithm based on Reinforcement Learning,” IEEE Transactions on Network Science and Engineering, doi: 10.1109/TNSE.2020.2995863, pp. 1-1, 2020.

\bibitem{j3}
J. Du, C. Jiang, H. Zhang, Y. Ren and M. Guizani, “Auction Design and Analysis for SDN-based Traffic Offloading in Hybrid Satellite-Terrestrial Networks,” IEEE Journal on Selected Areas in Communications, vol. 36, no. 10, pp. 2202-2217, 2018.

\bibitem{z3}
P. Zhang, C. Wang, Z. Qin and H. Cao, “A multidomain virtual network embedding algorithm based on multiobjective optimization for Internet of Drones architecture in Industry 4.0,” Software: Practice \& Experience, vol. 2020(SI), pp. 1-19, 2020.

\bibitem{z7}
P. Zhang, X. Pang, Y. Bi, H. Yao, H. Pan and N. Kumar, “DSCD: Delay Sensitive Cross-Domain Virtual Network Embedding Algorithm,” IEEE Transactions on Network Science and Engineering, doi: 10.1109/TNSE.2020.3005570, pp. 1-1, 2020.

\bibitem{b7}
R. Geng and H. Lu, ``Virtual Network Embedding Algorithm for Multi-domain SDN Networks,'' Journal of Chinese Computer Systems, vol. 37, no. 12, pp. 2593-2597, 2016.

\bibitem{b8}
L. Peng, ``A Multi-Domain Virtual Network Embedding Algorithm Based on Minimum Cost,'' Journal of South China University of Technology, vol. 43, no. 9, pp. 67-73+112, 2015.

\end{thebibliography}
\end{document}